\documentclass[fleqn,usenatbib]{mnras}

\usepackage{newtxtext,newtxmath}
\usepackage{soul}
\usepackage{multirow}

\usepackage[T1]{fontenc}

\DeclareRobustCommand{\VAN}[3]{#2}
\let\VANthebibliography\thebibliography
\def\thebibliography{\DeclareRobustCommand{\VAN}[3]{##3}\VANthebibliography}

\usepackage{graphicx}	
\usepackage{amsmath}	
\usepackage[table]{xcolor}

\title[Gas absorption towards $\eta$ Tel]{Gas absorption towards the $\eta$ Tel debris disc: winds or clouds?}

\author[D. P. Iglesias et al.]{
Daniela P. Iglesias,$^{1}$\thanks{E-mail: D.P.Iglesias@leeds.ac.uk (DPI)}
Olja Pani\'c,$^{1}$
and Isabel Rebollido$^{2}$
\\
$^{1}$School of Physics and Astronomy, Sir William Henry Bragg Building, University of Leeds, Leeds LS2 9JT, UK\\
$^{2}$Centro de Astrobiolog\'ia (CAB CSIC-INTA) ESAC Campus Camino Bajo del Castillo, s/n, Villanueva de la Cañada, 28692, Madrid, Spain\\
}

\date{Accepted XXX. Received YYY; in original form ZZZ}

\pubyear{2015}

\begin{document}
\label{firstpage}
\pagerange{\pageref{firstpage}--\pageref{lastpage}}
\maketitle

\newcommand{\isa}{\color{red}}
\definecolor{light-gray}{gray}{0.95}

\begin{abstract}$\eta$ Telescopii is a $\sim$23 Myr old A-type star surrounded by an edge-on debris disc hypothesised to harbour gas. Recent analysis of far- and near-ultraviolet spectroscopic observations of $\eta$ Tel  found absorption features at $\sim$-23 km\,s$^{-1}$ and $\sim$-18 km\,s$^{-1}$ in several atomic lines, attributed to circumstellar and interstellar gas, respectively. In this work, we put the circumstellar origin of the gas to a test by analysing high resolution optical spectroscopy of $\eta$ Tel and of three other stars with a similar line of sight as $\eta$ Tel: HD\,181327, HD\,180575, and $\rho$ Tel. We found absorption features at $\sim$-23 km\,s$^{-1}$ and $\sim$-18 km\,s$^{-1}$ in the Ca\,{\sc ii} H\&K lines, and at $\sim$-23 km\,s$^{-1}$ in the Na\,{\sc i} D1\&D2 doublet in $\eta$ Tel, in agreement with previous findings in the ultraviolet. However, we also found absorption features at $\sim$-23 km\,s$^{-1}$ in the Ca\,{\sc ii} K lines of the three other stars analysed. This strongly implies that the absorption lines previously attributed to circumstellar gas are more likely due to an interstellar cloud traversing the line of sight of $\eta$ Tel instead.

\end{abstract}




\section{Introduction}


The presence of gas in debris discs was thought to be rare and debris discs were considered as gas free dusty discs in previous decades. Nowadays, gaseous debris discs are widely accepted as the number of gas detections has increased along different wavelength ranges. The earliest detection was reported in \cite{Slettebak1975} around the star $\beta$ Pic, where sharp, deep absorption features are observed in the Ca\,{\sc ii} K and H lines. However, the circumstellar nature of these absorption lines was not attributed until \cite{Hobbs1985}. This gas, close to the star, is expected to be at temperatures higher than the sublimation temperature of refractory elements (i.e. >1000K), and affected by stellar winds. Given that such conditions would prevent the gas to be gravitationally stable in that region, replenishment mechanisms, such as exocomets have been invoked to explain this phenomena \citep{Ferlet1987,Beust1998}. Gas in debris discs can also be detected in emission mainly in the far-IR and millimeter wavelengths. Typical emission detections of gas-phase species are from CO, O, C and H$_{2}$, and they are mainly detected in debris discs that have higher dust masses and are young, within an age range $\sim$10-50 Myrs (e.g. \citealt{Kospal2013, Moor2017, Moor2019, Lieman-Sifry2016}). 

Gas emission detections are straight forward as the gas is being directly observed. However, absorption detections are indirect as the gas can only be observed in the stellar spectrum while it transits in front of the star. This makes the assessment of the circumstellar nature of the gas a bit more complicated, as absorption features in the stellar spectrum can also arise from traversing interstellar clouds or due to stellar variability such as pulsations or multiplicity. Studies such as \cite{Hales2017} and \cite{Iglesias2018} have focused in determining whether the absorptions seen in the spectra of debris disc hosting stars are of circumstellar or interstellar origin by studying the spectra of additional stars with a similar line of sight. The difficulty of determining the origin of gas absorption features can easily lead to misinterpretation of the observations. For instance, the case of HR\,10, a disc star observed to have variable absorption features in the Ca\,{\sc ii} K line, was initially interpreted to be transited by exocomets releasing gas as they orbit the star \citep{LagrangeHenri1990a, Welsh1998, Redfield2007b}. Later, \cite{Montesinos2019} showed that HR\,10 was actually a binary system, each star having an individual gaseous envelope. This generates two narrow absorptions lines that have variable radial velocities, as the binaries orbit each other. More recently, \cite{Eiroa2021} have reported that the star $\phi$ Leo, with a high rate of variations in its spectra \citep[attributed to exocomets years before by the same authors;][]{Eiroa16}, is a rapidly rotating $\delta$ Scuti star surrounded by a circumstellar disc, possibly supplied by the stellar pulsations. Therefore, further revision of absorption detections including additional information is sometimes necessary to confirm the origin of the gas.


Currently, the number of stars with detections of circumstellar gas in absorption (either metastable and -usually- centered at the radial velocity of the star or variable in flux and/or radial velocity) is over 30, being most of them debris discs stars that also show extended CO emission \citep[mostly observed with ALMA, e.g.][]{Lieman-Sifry2016, Moor2019}. However, the effects and dynamics of this gas in debris discs are not fully understood yet. A recent study by \cite{Kral2023} suggests that gas outflows or winds could be present in discs with low gas content, and cite the example of NO Lup \citep{Lovell2021}.
Up until now, two debris discs have been reported to show features consistent with winds in their spectra: $\sigma$ Her \citep{Chen&Jura2003} and $\eta$ Tel \citep{Rebollido2018, Youngblood2021}.

\begin{table*}
	\centering
	\caption{Properties of $\eta$ Tel and the comparison stars. Coordinates and spectral types are taken from Simbad database \citep{Wenger2000}. Number of observations used per instrument are indicated: HARPS=`H', UVES=`U', and FEROS=`F'. Angular separations $\theta$ are calculated with respect to $\eta$ Tel. Distances are from Gaia DR3 \citep{Gaia2022k,Gaia2016}.}
	\label{tab:nearby}
	\begin{tabular}{lccccccc} 
		\hline
		Name & RA & DEC & Sp. Type & rv$_{\star}$ & Number of & $\theta$ & Distance \\
          & [J2000] & [J2000] &  & & Observations & [degrees] & [pc] \\
          \hline
          $\eta$ Tel (or HD\,181296) & 19:22:51.21 & -54:25:26.15 & A0V & -0.55$\pm$1.56 & 39H, 29F & 0.00 & 48.54$\pm$0.23 \\
		HD\,181327 & 19:22:58.94 & -54:32:16.97 & F6 & -0.01$\pm$0.32 & 64H, 3U, 27F & 0.12 & 47.78$\pm$0.07 \\
		HD\,180575 & 19:19:37.42 & -51:34:21.37 & A1V & -28.75$\pm$2.21 & 1F & 2.89 & 144.20$\pm$0.72 \\
		  $\rho$ Tel (or HD\,177171) & 19:06:19.96 & -52:20:27.28 & F6 & -86$\pm$10 -- 93$\pm$15 & 39H & 3.23 & 58.32$\pm$0.33 \\
		\hline
	\end{tabular}
\end{table*}

\begin{figure*}
	\includegraphics[width=\textwidth]{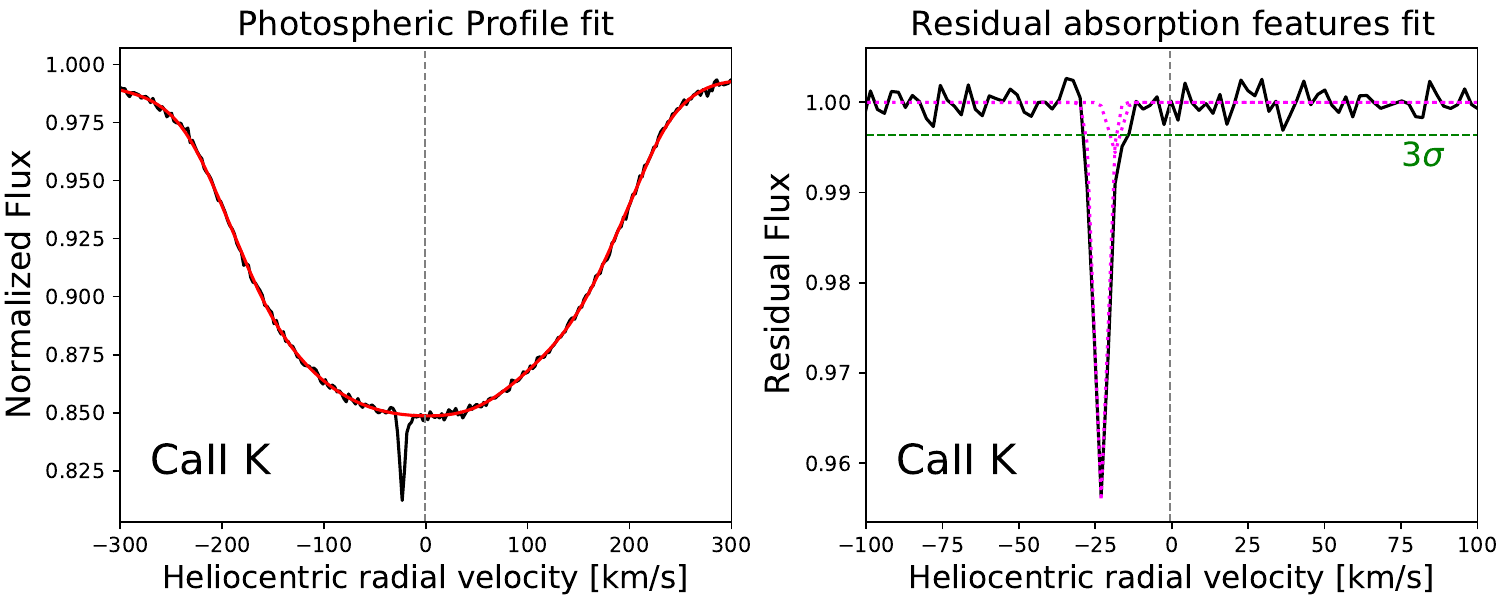}
    \caption{Profile fit of the Ca\,{\sc ii}\,K photospheric line of $\eta$ Tel (left side, in red), and residual absorption fit after normalizing by the photospheric fit (right side, in dotted pink line). Heliocentric radial velocity of $\eta$ Tel is marked with a gray dashed vertical line and the 3$\sigma$ level is marked in a green dashed horizontal line.
    }
    \label{fig:splineCaK}
\end{figure*}

In this paper we present evidence of matching absorption features seen in stars with a similar line of sight as $\eta$ Tel, strongly supporting an alternative explanation: a line of sight absorption by interstellar material unassociated to the star.

\section{Spectroscopic Observations}

We use 68 high-resolution (R~$>$~40,000) optical spectra of $\eta$ Tel obtained from the ESO archive\footnote{\url{http://archive.eso.org}}. A total of 77 spectra were available: 39 from HARPS (R$\sim$115000, $\lambda$$\sim$378nm-691nm; \citealt{Mayor2003}) and 38 from FEROS (R$\sim$48000, $\lambda$$\sim$350nm-920nm; \citealt{Kaufer1999}). The FEROS spectra observed in 2003 were discarded due to their low signal-to-noise ratio (SNR), leaving a total of 29. In addition, we searched for high-resolution spectroscopic observations of stars within a similar line of sight to $\eta$ Tel in the ESO archive. Any cloud that might be traversing the line of sight of $\eta$ Tel should be somewhere at an (unknown) distance between the observer and $\eta$ Tel, therefore the upper limit to the distance of such cloud is that of $\eta$ Tel. Considering that $\eta$ Tel is at a distance of 48.54pc \citep{Gaia2022k,Gaia2016}, we chose to search within a search-box of 8$^{\circ}$ width centered in the target, which would cover an area projected in the plane of the sky of width up to $\sim$6pc at such distance. Typical interstellar cloud sizes are in the order of $\sim$10pc \citep{Redfield08}, therefore, a projected area of width $\sim$6pc should be able to reasonably trace interstellar clouds in the line of sight of $\eta$ Tel. Ideally, the comparison targets should be at a similar distance (from Earth) to that of $\eta$ Tel for their light to be absorbed by the same interstellar clouds. Targets at larger distances are also useful to trace such clouds, however, their lines of sight might also be traversed by additional clouds that are further away than our science target.

The best targets to trace interstellar clouds are early-type stars (earlier than $\sim$F6) as they have fewer spectral lines and are often rotationally broadened, which makes it easier to resolve interstellar narrow absorption features. We found spectroscopic observations in the ESO archive of three early-type stars, listed in Table \ref{tab:nearby}, within the search box around $\eta$ Tel and we used them to probe traversing interstellar clouds in the line of sight. From now on, we will refer to them as `comparison stars'. Spectra from UVES (flexible resolving power and wavelength coverage; R$\sim$40000-110000, $\lambda\sim$300-1100nm: \citealt{Dekker2000}), HARPS, and FEROS were found for these stars, as listed in Table~\ref{tab:nearby}.

Observations were reduced with the corresponding instrument Data Reduction Software, and we applied heliocentric radial velocity corrections to the UVES spectra (this correction is already included in the pipelines of FEROS and HARPS). A posteriori, we corrected for telluric contamination using \texttt{MOLECFIT}\footnote{\url{http://www.eso.org/sci/software/pipelines/skytools/molecfit}} (\citealt{Smette2015}, \citealt{Kausch2015}). This tool models the absorption from the Earth’s atmosphere including molecules such as H$_{2}$O, O$_{2}$, and CO$_{2}$ (among others). These corrections are particularly important for the portion of the spectrum $>\sim$500nm.

\section{Analysis of the spectra}
\label{sec:analysis} 

\subsection{Absorption features}

We searched for narrow absorption features superimposed on the photospheric spectral lines of $\eta$ Tel that may indicate the presence of gas crossing the line of sight of the star, either due to winds, exocomets, a gaseous disc or interstellar clouds. We analysed several gas tracers such as Ca\,{\sc ii}, Mg\,{\sc ii}, Fe\,{\sc i}, Fe\,{\sc ii}, Ti\,{\sc ii}, Na\,{\sc i}, He\,{\sc i} and Balmer lines. However, we only found detectable absorption features in the Ca\,{\sc ii} H\&K lines at 3968.5\AA~ and 3933.7\AA, and in the Na\,{\sc i} D1\&D2 doublet at 5895.9\AA~ and 5889.9\AA, respectively.

\begin{figure*}
	\includegraphics[width=0.245\textwidth]{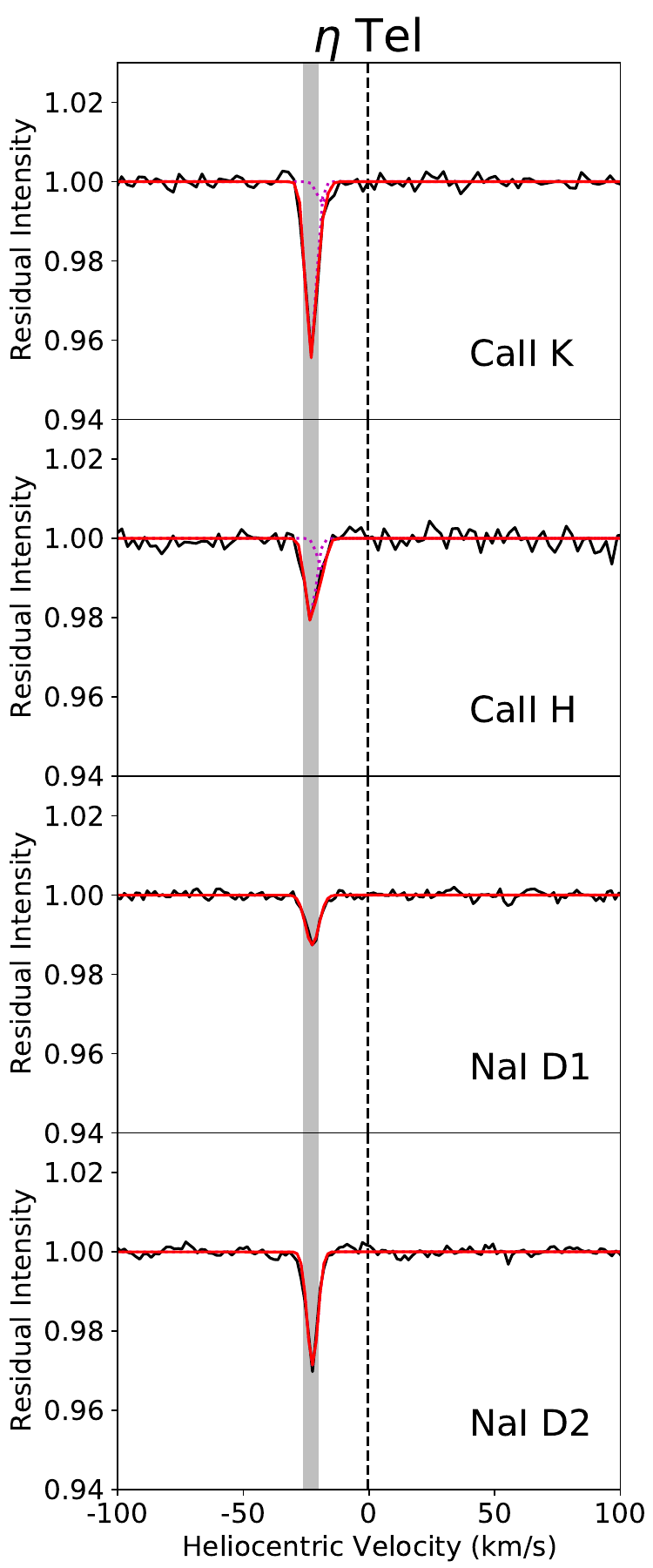}
 	\includegraphics[width=0.245\textwidth]{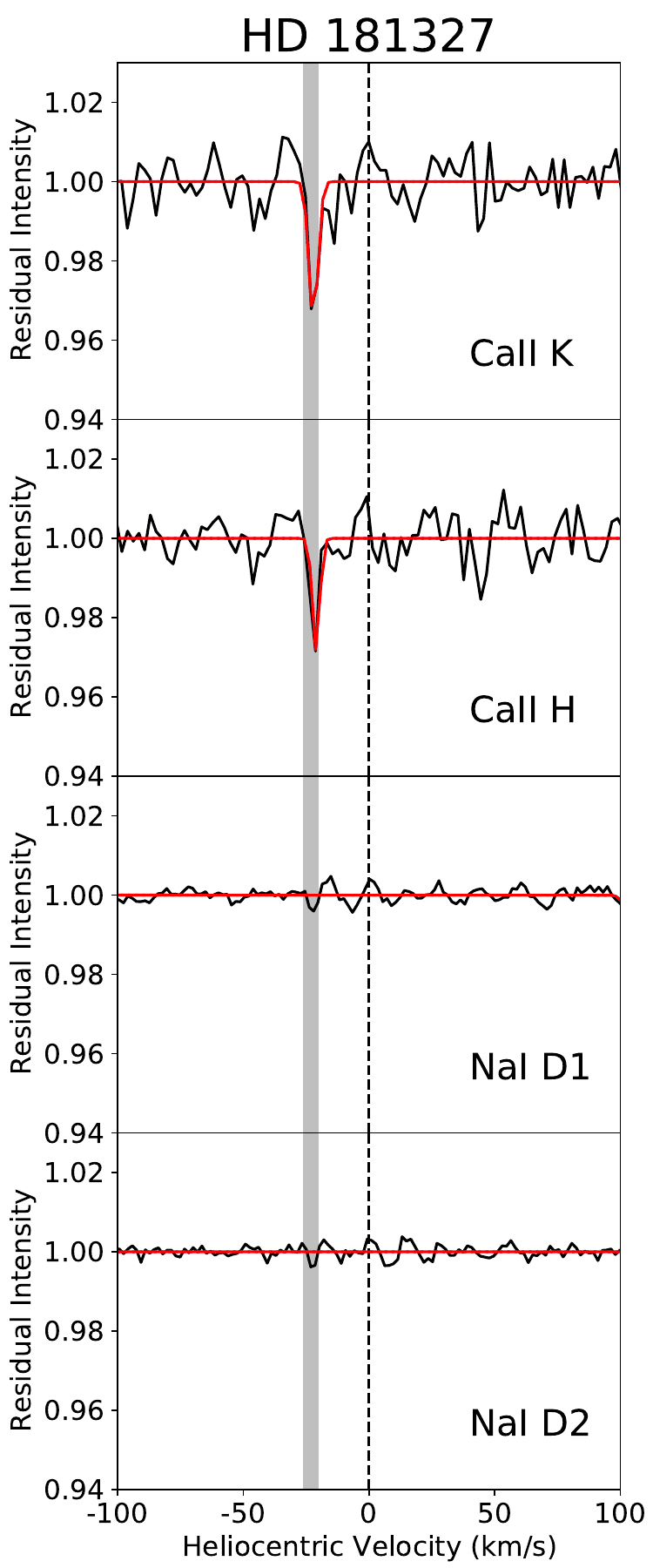}
   	\includegraphics[width=0.245\textwidth]{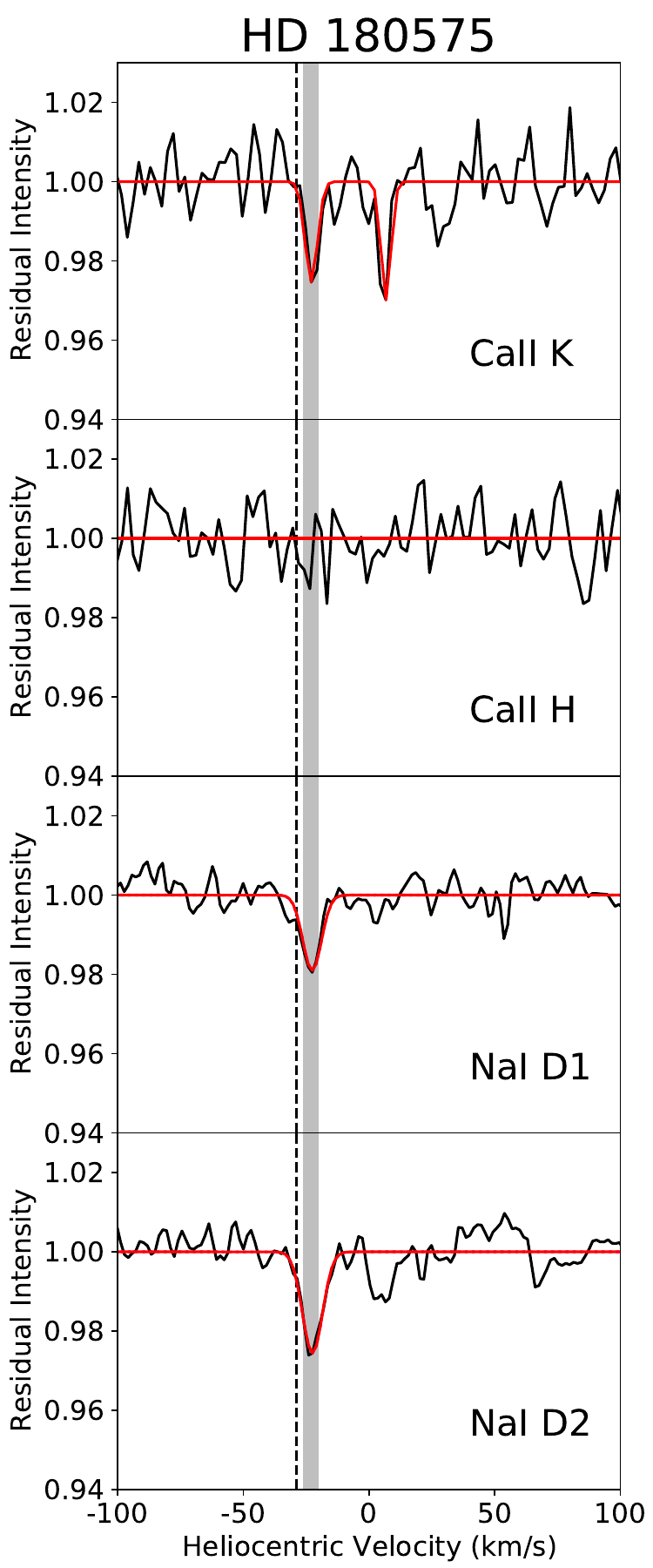}
    \includegraphics[width=0.245\textwidth]{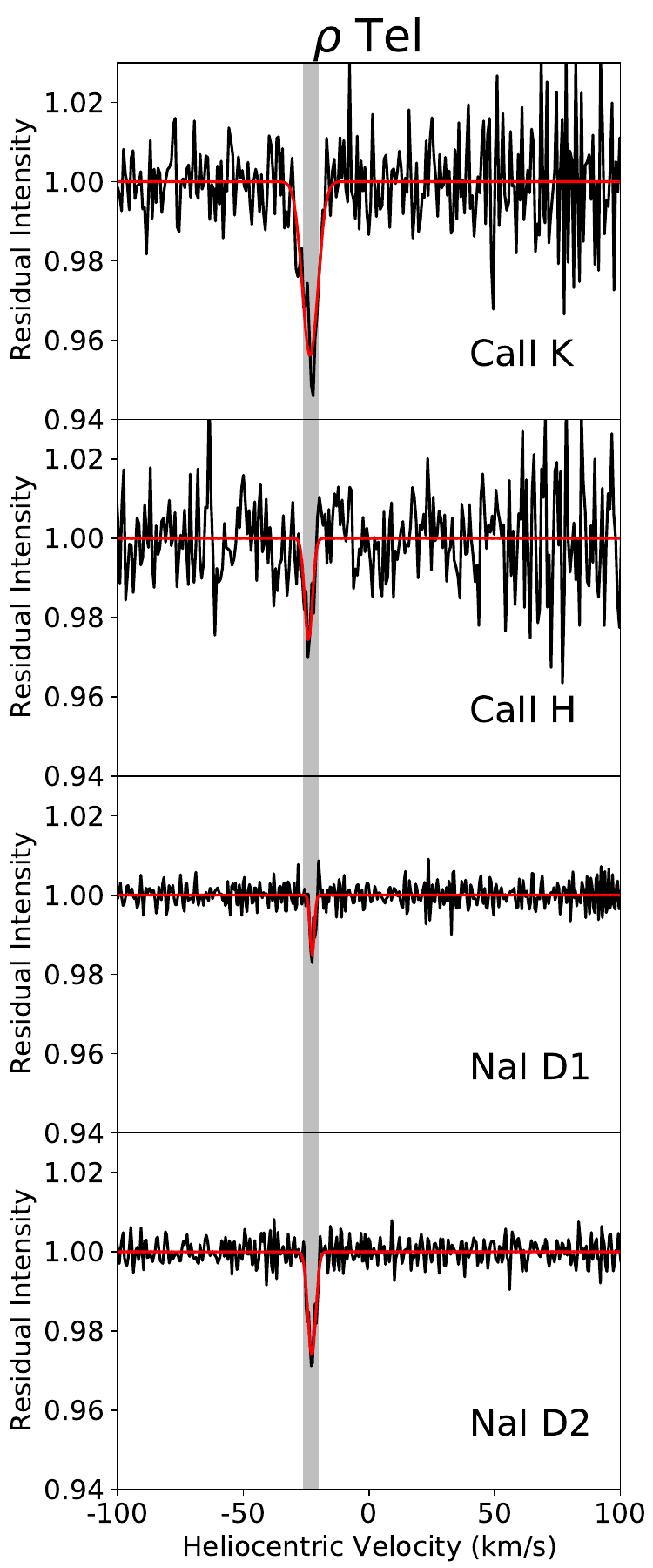}
    \caption{Detected absorption features in the lines of each star are shown with the red line, representing the best fit profile. For the case of $\eta$ Tel, the two individual Gaussian profiles are marked with a pink dotted line in the Ca\,{\sc ii} H\&K doublet. Heliocentric radial velocity for each star is marked with a dashed vertical line, except for $\rho$ Tel due to it being widely variable. The radial velocity of the absorption feature detected in the Ca\,{\sc ii} K line of $\eta$ Tel rv$_{\rm K}$=-23.10$\pm$3.14 km\,s$^{-1}$ is highlighted across all the figures in a gray line of width equal to the uncertainty.
    }
    \label{fig:4lines}
\end{figure*}

\begin{table*}
	\centering
	\caption{Properties of the absorption features detected in $\eta$ Tel and in the comparison stars. Radial velocity (rv), standard deviation of the pseudo continuum ($\sigma$), and intensity of the line in terms of sigma detection (Int.) are shown for each line.}
	\label{tab:feat}
	\begin{tabular}{|l|ccc|ccc|ccc|ccc|}
 \hline 
 & \multicolumn{3}{c|}{Ca\,{\sc ii} K } & \multicolumn{3}{c|}{Ca\,{\sc ii} H } & \multicolumn{3}{c|}{Na\,{\sc i} D1 } & \multicolumn{3}{c|}{Na\,{\sc i} D2 } \\

		Name & rv & $\sigma$ & Int. & rv & $\sigma$ & Int. & rv & $\sigma$ & Int. & rv & $\sigma$ & Int. \\
          & (km\,s$^{-1}$) &  & ($\sigma$) & (km\,s$^{-1}$) & & ($\sigma$) & (km\,s$^{-1}$) & & ($\sigma$) & (km\,s$^{-1}$) & & ($\sigma$) \\
          \hline
  \multirow{ 2}{*}{$\eta$ Tel} & -23.10$\pm$3.14 & 0.0012 & 36.71 & -23.27$\pm$3.10 & 0.0013 & 15.43 & -22.58$\pm$3.09 & 0.00087 & 14.51 & -22.45$\pm$2.63 & 0.0010 & 27.63 \\
             & -18.52$\pm$3.02 & 0.0012 & 4.63 & -18.74$\pm$2.91 & 0.0013 & 6.00 & -- & 0.00087 & -- & -- & 0.0010 & -- \\

        \rowcolor{light-gray}     HD\,181327 & -22.07$\pm$2.93 & 0.0031 & 11.45 & -21.00$\pm$2.70 & 0.0021 & 13.49 & -- & 0.0015 & -- & -- & 0.0014 & -- \\
		\multirow{ 2}{*}{HD\,180575} & -22.94$\pm$3.37 & 0.0061 & 4.19 & -- & 0.0053 & -- & -22.58$\pm$3.99 & 0.0035 & 5.48 & -22.45$\pm$4.28 & 0.0073 & 3.51 \\
            & 6.78$\pm$3.03 & 0.0061 & 4.93 & -- & 0.0053 & -- & -- & 0.0035 & -- & -- & 0.0073 & -- \\
		\rowcolor{light-gray}  $\rho$ Tel & -23.41$\pm$3.22 & 0.0076 & 5.80 & -24.02$\pm$1.70 & 0.0053 & 4.91 & -22.58$\pm$0.98 & 0.0026 & 5.98 & -22.74$\pm$1.64 & 0.0026 & 9.99 \\
		\hline
	\end{tabular}
\end{table*}

For the analysis, each observation of a photospheric line was normalized to the continuum and then a median spectrum was calculated from all the epochs. For the cases where observations had different resolutions, the observations were downgraded to match that of the lowest resolution and resampled to match the same wavelength grid in order to compute the median spectrum for each data point. This way, the SNR of the resulting spectrum is much higher than that of individual observations. Any non-photospheric stable feature should persist in this median spectrum. Then, a cubic spline function was fit to each photospheric line and used to normalize the line so that only residual noise and superimposed absorption features remain, as shown in Fig.~\ref{fig:splineCaK}. Any absorption having an intensity $>3\sigma$ in this `flat' residual spectrum is considered a detection. 

\begin{figure*}
	\includegraphics[width=0.547\textwidth]{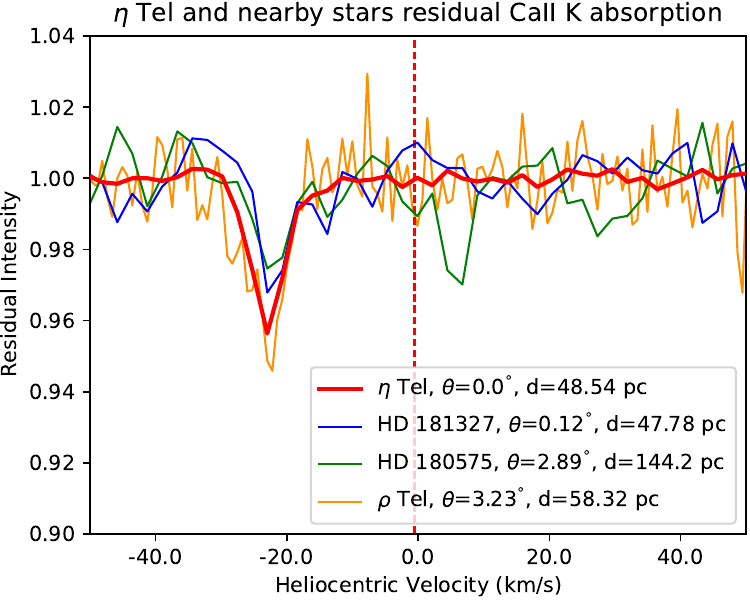}
 	\includegraphics[width=0.445\textwidth]{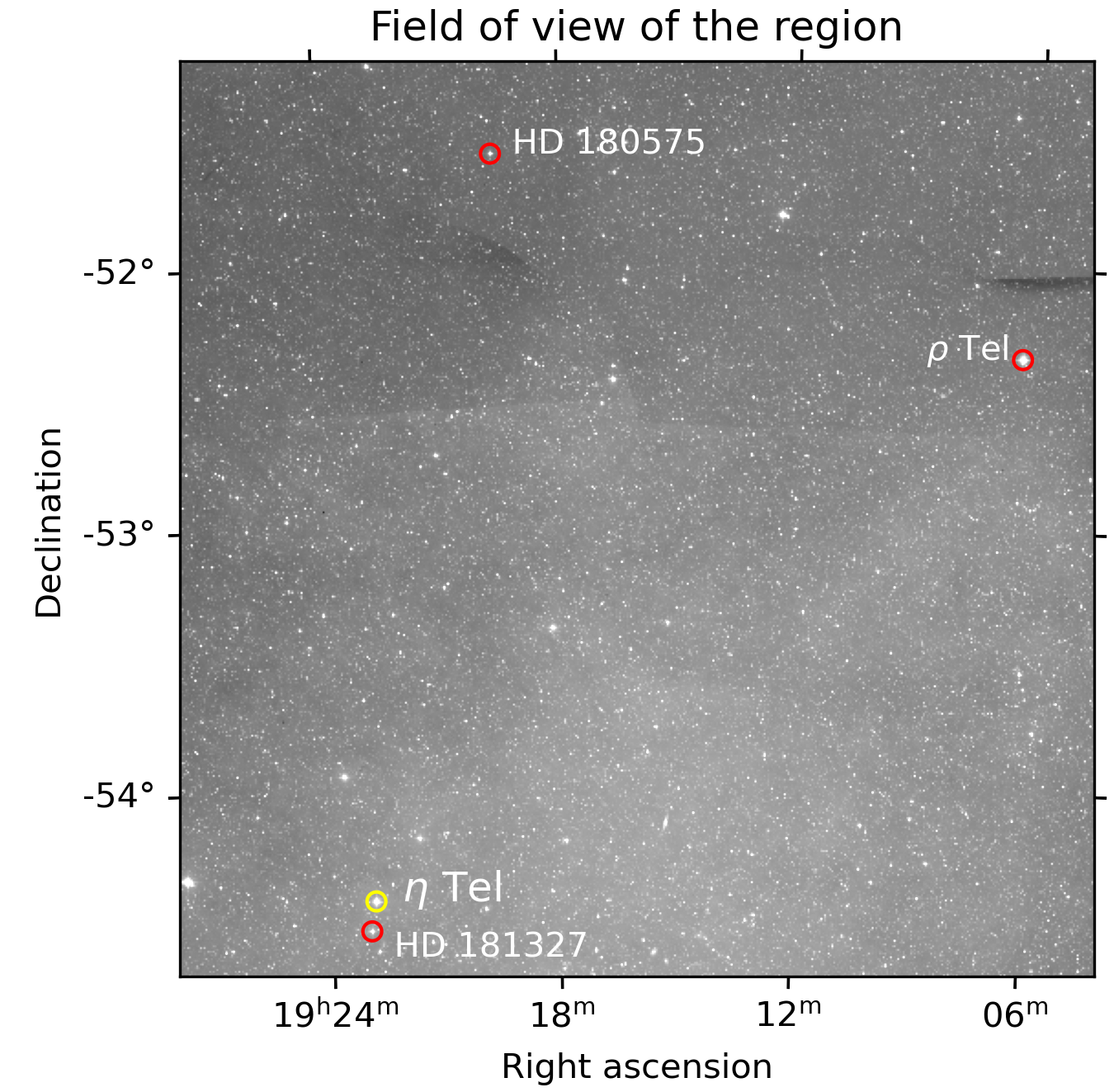}
    \caption{\textbf{Left:} Ca\,{\sc ii}\,K interstellar absorption present in stars with a similar line of sight as $\eta$ Tel. Heliocentric radial velocity of $\eta$ Tel is marked with a dashed red vertical line. \textbf{Right:} Position in the sky of the four targets. 
    }
    \label{fig:example_figure}
\end{figure*}

\begin{table}
	\centering
	\caption{Apparent column density $N$ and $\frac{N(Na\,I)}{N(Ca\,II)}$ ratio for the component detected at $\sim$-23 km\,s$^{-1}$ towards $\eta$ Tel and the comparison stars. Upper limit is given for the case of non detection of the Na\,{\sc i} D1\&D2 lines.}
	\label{tab:Density}
		\begin{tabular}{|l|ccc|}
 \hline 
 
		Name & N(Ca\,{\sc ii}) & N(Na\,{\sc i}) & \multirow{ 2}{*}{$\frac{N(Na\,I)}{N(Ca\,II)}$}  \\
        &  (cm$^{-2}$) &  (cm$^{-2}$)  & \\
          \hline
$\eta$ Tel & (3.6$\pm0.1)\times 10^{10}$  & (1.53$\pm0.06)\times 10^{10}$ & 0.43$\pm$0.02 \\

        \rowcolor{light-gray} HD\,181327  & (2.5$\pm0.2)\times 10^{10}$  & $<7.8\times 10^{8}$  & $<$0.03 \\
		HD\,180575  & (2.4$\pm0.6)\times 10^{10}$ & (2.6$\pm0.7)\times 10^{10}$ & 1.1$\pm$0.4 \\
		\rowcolor{light-gray}  $\rho$ Tel  & (5.3$\pm0.9)\times 10^{10}$ & (1.0$\pm0.1)\times 10^{10}$ & 0.19$\pm$0.04 \\
		\hline
	\end{tabular}
\end{table}

A Gaussian profile was fit to each absorption feature in order to measure its radial velocity and intensity. 
For the case of $\eta$ Tel, two profiles were used to fit the absorptions found in the Ca\,{\sc ii} H\&K doublet as the absorptions are blended (see Fig. \ref{fig:4lines}, first and second panel of $\eta$ Tel, or Fig. \ref{fig:splineCaK}, right side). The same analysis was performed for the three comparison stars. It is worth noting that $\rho$ Tel is a double-lined spectroscopic binary (SB2) \citep{Nordstrom2004} and each observation shifted significantly in radial velocity. However, after calculation of the median spectrum, a stable absorption feature (i.e. with constant flux) remained in the Ca\,{\sc ii} H\&K and Na\,{\sc i} D1\&D2 lines of $\rho$ Tel at a constant velocity, despite the radial velocity variations of the photospheric lines.

We also measured the apparent column densities $N$ for the features of main interest following \cite{Somerville1988}. For this method we used the oscillator strength values taken from \cite{Morton1991} and we used equivalent width $EW$ values that we calculated from our Gaussian profiles. Upper limits for the case of non detections were calculated by integrating over the pseudo-continuum. Uncertainties for $N$ were propagated from the $EW$ uncertainties which, in turn, were derived from the SNR and Gaussian profile goodness of fit. As a double-check, we also calculated $N$ following \cite{Savage1991} and the values obtained closely matched those using \cite{Somerville1988} methodology.

\subsection{Stellar radial velocity measurement}

Stellar radial velocities were calculated to test whether the absorption features were correlated to the radial velocities of the stars. We measured the radial velocity of these stars following the methodology of \cite{Iglesias2023} for early-type stars. We measured the centroid of their Balmer lines by fitting Lorentzian profiles and averaged the results for all the lines and epochs, taking the standard deviations as uncertainties. For the case of $\rho$ Tel, as it is an SB2, we fit two Lorentzian profiles and we provide the radial velocity range covered by the pair. The stellar radial velocity measurements (rv$_{\star}$) are given in Table~\ref{tab:nearby}.

\section{Results and Discussion}

\subsection{Stellar radial velocity}

We obtained a stellar heliocentric radial velocity rv$_{\star}$=-0.55$\pm$1.56 km\,s$^{-1}$ for $\eta$ Tel. This is in agreement with some values reported in the literature such as rv$_{\star}$=-1.290$\pm$0.0456 km\,s$^{-1}$ \citep{Zuñiga2021} and rv$_{\star}$=-3.0$\pm$3.0 km\,s$^{-1}$ \citep{Rebollido2018}, and differ from others; e.g. rv$_{\star}$=-5.6$\pm$2.8 km\,s$^{-1}$ \citep{Youngblood2021}, rv$_{\star}$=-5.0$\pm$1.5 km\,s$^{-1}$ \citep{Rebollido2020}, and rv$_{\star}$=+13.0$\pm$3.7  km\,s$^{-1}$ \citep{Kharchenko2007}. The spread in measurements found in the literature is explained by the difficulty to obtain precise radial velocities for A-type stars given their few and rotationally broadened spectral lines. For HD\,181327, values in the literature range between rv$_{\star}$=-0.690$\pm$0.6245  km\,s$^{-1}$ \citep{Zuñiga2021} and rv$_{\star}$=0.10$\pm$0.40  km\,s$^{-1}$ \citep{Gontcharov2006}. We calculated rv$_{\star}$=-0.01$\pm$0.32  km\,s$^{-1}$, which is consistent with previous measurements. For HD\,180575, we obtained rv$_{\star}$=-28.75$\pm$2.21  km\,s$^{-1}$, which is consistent with the only measurement found in the literature rv$_{\star}$=-30.23$\pm$0.95  km\,s$^{-1}$ \citep{Gaia2022k}. Finally, for $\rho$ Tel being an SB2, we found that the binary components shift in rv$_{\star}$ from $\sim$-86 to $\sim$93 km\,s$^{-1}$. It is beyond the scope of this paper to obtain the full orbital parameters of the binary pair, but these values are consistent with the mean radial velocity of the pair and its standard deviation found in the literature; rv$_{\star}$=0.1$\pm$51.5  km\,s$^{-1}$ \citep{Nordstrom2004}. 

\subsection{Detected absorption features}

The absorption features detected in each object are shown in Fig.~\ref{fig:4lines} and their parameters are summarized in Table~\ref{tab:feat}.
We identified two absorption features in the Ca\,{\sc ii} H\&K lines at $\sim$-23 and $\sim$-18 km\,s$^{-1}$ and one in the Na\,{\sc i} D1\&D2 doublet at $\sim$-23 km\,s$^{-1}$ in $\eta$ Tel. One absorption feature was detected in the Ca\,{\sc ii} H\&K lines in HD\,181327 at $\sim$-22 km\,s$^{-1}$. Two absorption features were also detected in the Ca\,{\sc ii} K line at $\sim$-23 and $\sim$7 km\,s$^{-1}$ and one in the Na\,{\sc i} D1\&D2 lines at $\sim$-23 km\,s$^{-1}$ in HD\,180575. Finally, one absorption feature was detected in the Ca\,{\sc ii} H\&K, and Na\,{\sc i} D1\&D2 doublets in $\rho$ Tel at $\sim$-23 km\,s$^{-1}$. 

The radial velocity of the absorption features found in the Ca\,{\sc ii} K line of the three comparison stars matches that of the absorption line found at $\sim$-23 km\,s$^{-1}$ in $\eta$ Tel. This can be seen in Fig.~\ref{fig:4lines} where we highlight the radial velocity of the absorption feature detected in the Ca\,{\sc ii} K line of $\eta$ Tel rv$_{\rm K}$=-23.10$\pm$3.14 km\,s$^{-1}$ in all the targets, for ease of comparison. We observe no correlation between the velocity of this absorption feature and the radial velocity of the stars. The feature is always at the same velocity independently of being redshifted or blueshifted with respect to the stars. Furthermore, in the case of $\rho$ Tel having SB2 components of highly variable radial velocity, the feature remains at a constant velocity showing no correlation to the binary pair. This non-photospheric absorption component at $\sim$-23 km\,s$^{-1}$ in the Ca\,{\sc ii} K line of $\eta$ Tel was previously identified by \cite{Welsh2018} where its origin was associated to the local interstellar cloud (LIC) complex. It was also reported by \cite{Rebollido2018}, but in this case, it was attributed to circumstellar origin. Furthermore, \cite{Youngblood2021} detected absorption lines at $\sim$-23 km\,s$^{-1}$ in C\,{\sc i}, C\,{\sc ii}, N\,{\sc i}, O\,{\sc i}, Mg\,{\sc ii}, Al\,{\sc ii}, Si\,{\sc ii}, S\,{\sc ii}, Mn\,{\sc ii}, and Fe\,{\sc ii} in $\eta$ Tel, which were attributed to radiatively driven winds due to being blue-shifted with respect to the radial velocity of the star and because some of the lines arise from exited energy levels not frequently observed in the local interstellar medium.

However, the fact that absorption lines at the corresponding velocity are also observed in three other stars with a similar line of sight as $\eta$ Tel means that the most plausible explanation is that these absorption features arise from an interstellar cloud, which traverses the line of sight to these stars. The three matching absorption features in the Ca\,{\sc ii} K line and the position of the stars in the sky can be seen in Fig.~\ref{fig:example_figure}.  In particular, HD\,181327 is at a very small angular separation with respect to $\eta$ Tel ($\theta$=0.12$^{\circ}$) and at a very similar distance. This strengthens the evidence of the absorptions coming from the same cloud as it discards the possibility of it coming from a cloud at a larger distance than $\eta$ Tel, for this comparison star. However, the non detection of Na\,{\sc i} absorption lines in HD\,181327 is puzzling as we would expect to detect them as well. We speculate that this might be due to its spectral type. HD\,181327, being of later F-type, has very strong photospheric Na\,{\sc i} lines, while $\eta$ Tel, an early A-type star has an almost flat, rotationally broadened Na\,{\sc i} doublet, where narrow absorption features are easily distinguishable. We suspect that (if any) Na\,{\sc i} absorptions in HD\,181327 might be weak and blended with the strong photospheric component. In the case of $\rho$ Tel, although also of F-type, the Na\,{\sc i} photospheric lines are much weaker in comparison to HD\,181327, and as they vary in radial velocity the stable narrow absorption feature was often detected in a shifted position with respect to the photospheric component. Another possibility is that the line of sight of HD\,181327 might be traversed by less material as it is slightly closer to Earth than $\eta$ Tel. The other two stars, HD\,180575 and $\rho$ Tel, have a larger angular separations ($\theta\sim$3$^{\circ}$), which, projected at the distance of $\eta$ Tel corresponds to $\sim$2pc. Interstellar clouds overall cover areas of width larger than 2pc, so it is very likely that the absorption in these other two stars comes from the same cloud as well. The fact that this absorption is observed in the four stars simultaneously makes it unlikely to be a coincidence and presents strong evidence favouring the origin to be an interstellar cloud traversing their line of sight, in agreement with the verdict by \cite{Welsh2018}.

In \cite{Youngblood2021}, the dynamics of the gas absorption lines in $\eta$ Tel was analyzed by computing the ratio $\beta$ of radiation pressure to gravitational force that atoms and molecules receive from the central star in the circumstellar gas scenario. They found no correlation between $\beta$ and the radial velocity of the absorptions. Furthermore, they mention that gas particles with $\beta<0.5$ should be bound to the star, while those with $\beta>0.5$ should be ejected from the system in hyperbolic orbits. However, the species detected in \cite{Youngblood2021} display a wide range of $\beta$ values ranging from 0.05 to 365.7, but they all are found at the same blueshifted radial velocity with respect to the star. 
An interstellar origin for these absorption features explains the fact that all the species are found at the same radial velocity regardless of each species having different radiation pressure coefficients. An interstellar origin is in agreement with the gas behaving as a single fluid instead of blowing out as a radiatively driven disc wind which would affect each species differently (i.e. their absorptions would be found at different radial velocities).

We also detect the feature at $\sim$-18 km\,s$^{-1}$ reported as interstellar by \cite{Youngblood2021} in the Ca\,{\sc ii} H\&K lines of $\eta$ Tel. Indeed, it matches the radial velocity of the G cloud towards $\eta$ Tel's sighline rv$_{\rm G}$=-18.51$\pm$1.51 km\,s$^{-1}$ \citep{Redfield08}. However, it is not detected in the other targets. This is likely due to it being very weak and the other targets having a lower SNR than $\eta$ Tel. We detect an additional absorption feature at $\sim$7 km\,s$^{-1}$ in the Ca\,{\sc ii} K line of HD\,180575 which is probably due to another cloud, as HD\,180575 is considerably more distant than the other stars and thus, likely to be traversed by further interstellar material.

Finally, we performed further analysis for the absorption features at $\sim$-23 km\,s$^{-1}$ to investigate whether their abundances were consistent with typical values of the interstellar medium or rather unusual. We calculated apparent column density values $N$ and the abundance ratio $\frac{N(Na\,I)}{N(Ca\,II)}$ for $\eta$ Tel and the comparison stars. These values are presented in Table \ref{tab:Density}. We compared with studies of the Na\,{\sc i} and Ca\,{\sc ii} abundances in the interstellar medium such as \cite{Crawford1992} and \cite{Welsh2010}. \cite{Crawford1992} shows a distribution of $\frac{N(Na\,I)}{N(Ca\,II)}$ ratios with respect to the local standard of rest (LSR) velocity of the components (see Fig.5 in their paper). In the LSR frame, our detected components at $\sim$-23 km\,s$^{-1}$ in the four targets would be within $\sim$-20 -- -19 km\,s$^{-1}$. At similar LSR velocities, the observed column density ratios range between $\sim$0.16--1.6, which is in agreement with our observed values ranging $\sim$0.19--1.1 (excluding the non detection of Na\,{\sc i} discussed above). On the other hand, \cite{Welsh2010} studied $\frac{N(Na\,I)}{N(Ca\,II)}$ ratios as a function of distance and concluded that for distances $<$80pc these ratios range between 0.1--1.0, while for larger distances they increase to values between 0.5--20. This, again, is in agreement with the observed ratios for our stars and their distances, being HD\,180575 the most distant and the one with the highest density ratio. This evidence, although it cannot fully confirm or rule out the origin of the gas by itself, further reassures the interstellar cloud scenario.

\section{Conclusions}

We analysed the high-resolution optical spectra of $\eta$ Tel available in the ESO archive and detected two non-photospheric absorption lines in the Ca\,{\sc ii} H\&K doublet at $\sim$-23 and $\sim$-18 km\,s$^{-1}$, and one in the Na\,{\sc i} D1\&D2 doublet at $\sim$-23 km\,s$^{-1}$. In addition, we performed the same analysis for three early-type stars with a similar line of sight as $\eta$ Tel: HD\,181327, HD\,180575, and $\rho$ Tel, and found a matching absorption feature at $\sim$-23 km\,s$^{-1}$ in the Ca\,{\sc ii} K line of the three stars, and matching absorptions in the Ca\,{\sc ii} H line and Na\,{\sc i} D1\&D2 doublet for at least two of these stars. 

This evidence strongly implies that the absorption feature found in several species in $\eta$ Tel in this and previous studies is due to an interstellar cloud in the line of sight of $\eta$ Tel and is unlikely to be of circumstellar origin. In addition, this would explain the gas acting as a single fluid instead of being blown out by radiatively driven winds as all species were found at the same radial velocity independently of their individual radiation pressure coefficients.

\section*{Acknowledgements}

D.P.I. and O.P. acknowledge support from the Science and Technology Facilities Council via grant number ST/T000287/1. I.R. is supported by grant FJC2021-047860-I and PID2021-127289NB-I00 financed by MCIN/AEI /10.13039/501100011033 and the European Union NextGenerationEU/PRTR. The authors thank Allison Youngblood for helpful discussions. The authors thank the anonymous referee for the useful feedback that helped improve the paper. This work has made use of data from the European Space Agency (ESA) mission Gaia (\url{https://www.cosmos.esa.int/gaia}), processed by the Gaia Data Processing and Analysis Consortium (DPAC, \url{https://www.cosmos.esa.int/web/gaia/dpac/consortium}). Funding for the DPAC has been provided by national institutions, in particular the institutions participating in the Gaia Multilateral Agreement. Based on observations collected at the European Southern Observatory under ESO programmes: 0101.A-9012(A), 0102.C-0584(A), 0103.C-0206(A), 073.C-0733(C), 073.C-0733(D), 075.A-9010(A), 076.A-9013(A), 077.C-0295(A), 077.C-0295(B), 077.C-0295(D), 078.C-0209(A), 079.A-9007(A), 079.A-9009(A), 079.C-0556(A), 080.C-0712(A), 083.C-0794(A), 083.C-0794(B), 083.C-0794(C), 083.C-0794(D), 084.C-1039(A), 085.A-9027(B), 088.C-0506(A), 093.A-9008(A), 094.A-9012(A), 098.C-0042(A), 098.C-0739(A), 099.A-9029(A), 099.C-0205(A), 179.C-0197(A), 192.C-0224(B), 192.C-0224(C).

\section*{Data Availability}

The data underlying this article are available in the ESO archive at \url{http://archive.eso.org/cms.html} and in the Gaia archive at \url{https://ge a.esac.esa.int/archive/}.



\bibliographystyle{mnras}
\bibliography{biblio} 




\appendix




\bsp	
\label{lastpage}
\end{document}